\documentstyle[eqsecnum,aps,amstex,epsfig,float,12pt]{revtex}

\def\ep{{\mbox{e}}}
\newcommand{\J}{{\mathrm{J}}}

\begin{document} 
\baselineskip = 24pt
\begin{center}
{\bf Compensation for thermal effects in  mirrors
of Gravitational Wave Interferometers.}\\
           {P. Hello}\\
Laboratoire de l'Acc\'el\'erateur Lin\'eaire\\
Universit\'e Paris-Sud - B\^atiment 208 - BP 34\\
 F-91898 Orsay cedex - France\\
\end{center}

Abstract : In this paper we study several means of compensating for thermal lensing
which, otherwise, should be a source of concern for future upgrades of interferometric
detectors of gravitational waves. The methods we develop are based on the principle
of heating the cold parts of the mirrors. We find that thermal compensation can help
a lot but can not do miracles. It seems finally that the best strategy for future upgrades
(``advanced configurations'') is maybe to use thermal compensation together with
another substrate materials than Silica, for example Sapphire. 

\baselineskip = 1.0\baselineskip

\section{Introduction}

Large scale interferometric detectors of gravitational waves such as LIGO\cite{ligo} or VIRGO\cite{virgo}
are currently under achievement. Their first data taking should begin in a few years from now.
But from their planned sensitivity and from what is currently known from astrophysical sources, 
direct detections of gravitational waves is far from sure to happen with this first generation of interferometers.
Plans for improving the sensitivity are already being developped, and the design of second generation interferometers
is well advanced \cite{ligo2}. Among the noises that have to be lowered, the shot noise is one of the
fundamental ones. It is indeed the dominant one in the high frequency region (typically above 1 kHz)
of the sensitivty of first generation interferometers.
The shot noise improvement does not amount to naively increase the power of the laser source for example,
since the mirrors of the interferometer have low but non zero absorptions. This results in thermal gradients
in the mirrors and a possible degradation of the detector performance.
Thermal effects in mirrors have been shown to be under control for the present design 
of interferometric detectors \cite{winkler,hel3} but they are likely to be one source of 
concern for future upgrades \cite{hel4} and even to introduce a new source of noise via the coupling
of laser power fluctuations to absorption asymmetry in the interferometer arms
\cite{hv}. Indeed, lowering the shot noise limited sensitivity by one
order of magnitude means increasing the circulating powers by two.
The planned
circulating powers in VIRGO for example are of the order of 15 kW stored in the kilometric 
cavities and about 1 kW crossing the (Silica) substrates of the cavity input mirrors.
The measured absorptions in the Silica susbtrates are about 1 ppm/cm for
the Heraeus 311SV \cite{lor}, and the best absorptions in the coatings have been measured
to be  about 1 ppm \cite{rempe,uehara,mackowski}.
With these figures, there will be
about 25 mW dissipated in the input mirrors (10 mW due to absorption
in the susbtrate and 15 mW due to absorption in the coating).
This gives reasonable losses (mainly due to the thermal lensing effect)
which does not affect too much the sensitivity. But, it is an other
story if the circulating powers increase by as low as a single factor 10,
for example if we keep the interferometer as it is, and simply
upgrade the laser to a 10 times more powerful one (upgrade from 10 W to 100 W in the case of VIRGO). 
In the current design of ``Advanced LIGO'' \cite{ligo2bis}, a laser source of about 100 W is planned
and hundreds of kW are planned to be stored on the
high finesse arm cavities (about 500 kW in case of Silica mirrors up to about 800 kW in case of Sapphire
mirrors).Clearly, at a time when R\&D is going to be funded for advanced configurations
of gravitational-wave detectors, means have to be investigated in order
to control the induced thermal lensing which otherwise could be desastrous.

The thermal lensing takes its source in the apparation of temperature gradients
in the mirrors, these gradients being  due to the non uniform  beam intensity
crossing the mirrors or hiting the coatings. In this article, we develop some means
to decrease
these gradients. The basic idea is very simple: we have to heat the ``cold''
parts of the mirrors in order to homogenize the temperature field. 
This can be done by an external light source with a 
proper profile, at a proper wavelength,
and being absorbed by the parts of the mirror that are not
heated by the main (about Gaussian) beam itself, or by a thermostat system
that fixes the external temperature of the mirrors to some value (principle
of the 'electric blanket'). Note that the implemantation of such a concept is planned
for ``Advanced LIGO'' \cite{ligo2bis}.
The goal of the article is of course not to provide a technical solution to 
this problem
but rather to indicate the prospects of the idea of compensating for
temperature gradients by heating the mirrors. First we will recall how we
estimate the temperature field and the induced thermal lens for a mirror
heated by a Gaussian beam either by absorption in its substrate or
by absorption in its coating and suspended in vacuo. We will then
give some orders of magnitude of the lensing effect and related losses for
first generation interferometer  and then for an upgrade. We will take the VIRGO numbers
for the numerical evaluations, but without really a loss of generality.
For the upgrade we will assume
simply a 100 W laser instead of a 10 W laser so all the circulating powers
have to be multiplied by 10, so 10 kW crossing the input mirrors and
150 kW stored inside the long cavities. In a second part, we will
consider the case of a mirror heated by an extra beam with a ring-shaped
profile as the one roughly produced by a battery of diodes located all around
the mirror (radial heating ogf the mirror). We will then consider the case of 
a beam coaxial with the main beam and again with a ring-shaped
profile (in this case we have to take into account the damping of this beam
due a priori to a strong absorption along the mirror axis). Finally we
will consider the case of a mirror with its lateral side ``in contact'' with
a thermostat of fixed temperature. In each case we will compute
the losses of the mirror, for a one way propagation of a perfect TEM$_{00}$
beam through the mirror (and so experiencing the thermal lensing), for
various absorbed powers from the external light source or various thermostat
temperatures (the main parameters of the studies), in order to look
for an optimum (minimal losses).
We will also consider briefly the case of Sapphire substrates, instead of Silica ones, but the substrate
materials choice will not change our conclusion. We will finally discuss the influence of the (main) beam size.

\section{Temperature and thermal lens in a mirror}

\subsection{ The mirror, the Heat equation and the boundary conditions}
We consider an long cavity input mirror: a Silica substrate
of width $h \simeq 10$ cm and radius $a\simeq 20$ cm with a reflective coating
located in the intra-cavity side. The mirror is heated by a 'laser' beam,
assumed in this part to be a perfect TEM$_{00}$ beam with a waist $w_0 \simeq 2$ cm
either in the substrate, where the circulating 
power is $P_{sub}$, or in the coating, where the power is $P_{coat}$. 
The coating, modelled as a very thin layer, is located at $z=-h/2$
(see fig.1).

We will note $\alpha$ the lineic absorption coefficient in the substrate
and $\epsilon$ the absorption coefficient in the coating. If alone in vacuo,
the mirror can only loose heat by radiation. Fortunately the temperature
rise in the mirror is supposed to be moderate so the radiative heat flux
can be linearized (this will be assumed in all the paper):
\begin{equation}
F=\sigma (T^4-T_e^4) \simeq 4 \sigma T_e^3 (T-T_e)
\label{flux}
\end{equation}
where $T_e$ is the temperature of the vacuum tank (supposed uniform)
and $\sigma$ is the Stefan constant (corrected for Silica emissivity).
In this part, starting from now, 
 we will note $T$ not for the absolute temperature, but its
deviation from $T_e$ ($T-T_e \longrightarrow T$); it is valid since 
everything is now linear.
The steady-state Heat equation to solve for the mirror is then:
\begin{equation}
- K \Delta T = \alpha I(r)
\label{heat}
\end{equation}
together with the boundary equations:
\begin{align}
- &K \frac{\partial T}{\partial r}(r=a, z) = 4e\sigma{T_0}^3 T(r=a, z) \label{cl1}\\
- &K \frac{\partial T}{\partial z}(r, z = h/2) = 4e\sigma{T_e}^3 T(r, z=h/2)\label{cl2}\\
- &K \frac{\partial T}{\partial z}(r, z =-h/2) = \epsilon I(r) - 4e\sigma{T_e}^3 T(r, z=-h/2)\label{cl3}
\end{align}
The linearisation of thermal equations (boundary conditions) implies that we can
study separatly the different sources of heating:  absorption by the coating
or the substrate of the Gaussian beam or of any other beam. Note that, in Eq.(\ref{heat}), 
we have
neglected the possible attenuation of the beam through the mirror. This is of course
valid since we consider here only the main laser beam for which the substrate is a low loss
one. We will have to take into account this effect for auxiliary beams if their wavelength
is within absorption bands of the Silica.

\subsection{Absorption in the coating}

If we consider only absorption in the coating, then the Heat equation (\ref{heat})
becomes the homogeneous one $\Delta T=0$, a general solution of which being expressed
as a Dini series:
\begin{equation}
T(r,z)= \sum_m \left( A_m \ep^{k_m z} + B_m \ep^{-k_m z}\right) \J_0 (k_m r) ,
\label{sol}
\end{equation}
which respects well the cylindrical symmetry of the problem.
The first boundary condition (\ref{cl1}) shows then that the coefficients 
$\zeta_m = a\times k_m$
are the zeros of the function
\begin{equation}
x  \J_1 (x) - \tau  \J_0 (x) = 0 ,
\end{equation}
where $\tau$ is the reduced radiation constant: $\tau = 4e\sigma T_e^3 a/K$.
The functions $\J_0 (k_m r)$ are then a complete set of orthogonal functions, for functions
defined in $[0,a]$, with scalar
products given by \cite{handbook}
\begin{equation}
\int_0^a \J_0 (k_m r) \J_0 (k_n r) r dr = \frac{1}{2} a^2 \left(1+ \frac{\tau^2}{\zeta_m^2}
\right) \J_0 (\zeta_m)^2  \delta_{nm}. \label{norm}
\end{equation}
The coefficients $A_m$ and $B_m$ are then given by the two last boundary
conditions (\ref{cl2}) and (\ref{cl3}), and we find finally the temperature:
\begin{equation}
T(r,z) = \sum_m \frac{\epsilon p_m a}{K} \ep^{-k_mh/2a} \,\,
\frac{(\zeta_m-\tau)\ep^{-k_m(h-z)}+(\zeta_m+\tau)\ep^{-k_mz}}
{(\zeta_m+\tau)^2-(\zeta_m-\tau)^2 \ep^{-2k_mh}} \,\, \J_0 (k_m r)
\end{equation}
where the $p_m$ are the coefficients of the Dini expansion of the intensity $I(r)$
\begin{equation}
I(r) = \sum_m p_m \J_0 (k_m r)
\end{equation}
The $p_m$ are then given by, taking account the normalisation condition
of Eq.(\ref{norm}):
\begin{equation}
p_m = \frac{2}{a^2}\frac{1}{1+\tau^2/\zeta_m^2}\frac{1}{\J_0 (\zeta_m)^2}
\int_0^a I(r) \J_0 (k_m r) r dr.
\label{pm}
\end{equation}
In this section we consider only the heating by the main beam (assumed to be
a perfect TEM$_{00}$ of power $P$) 
\begin{equation}
I(r) = \frac{2P}{\pi w_0^2} \ep^{-2r^2/w_0^2}
\end{equation}
so that $p_m$ can be computed as
\begin{equation}
p_m =\frac{P}{\pi a^2}\frac{1}{1+\tau^2/\zeta_m^2}\frac{1}{\J_0 (\zeta_m)^2}
\ep^{-k_m^2 w_0^2/8}
\end{equation}
at the condition $a >> w_0$ \cite{handbook2}.

The optical path distortion due to thermal lensing is then given by \cite{hel1}:
\begin{equation}
\psi(r) = \frac{dn}{dT} \int_{-h/2}^{+h/2} T(r,z) dz \label{lens}
\end{equation}
where $dn/dT$ is the thermal index coefficient of the substrate materials.
After straightforward integration, we find
\begin{equation}
\Psi(r) = \frac{dn}{dT}\frac{a}{K} \sum_m \frac{\epsilon p_m}{k_m}
\frac {1 - \ep^{-k_m h}}{(\zeta_m+\tau)-(\zeta_m-\tau)\ep^{-k_m h}} \, \J_0 (k_m r).
\label{coatlens}
\end{equation}

Some quantitative results and figures are given below.

\subsection{Absorption in the substrate}

In this case we have to solve the full Eq.(\ref{heat}). For this purpose, 
we add to the general solution Eq.(\ref{sol}) of the homogeneous equation
a particular solution. A particular solution of Eq.(\ref{heat}) is easily
found to be $\sum_m \frac{\alpha p_m}{K k_m^2} \J_0 (k_m r)$ so that the complete
solution is 
\begin{equation}
T(r,z) = \sum_m \left(  A_m \ep^{k_m z} + B_m \ep^{-k_m z}
+\frac{\alpha p_m}{K k_m^2} \right) \J_0 (k_m r)
\label{subsol}
\end{equation}
The boundary condition (\ref{cl1}) gives again that $k_m \times a$ is the m-th zero
of $x \J_1 (x)-\tau\J_0 (x)$ and the two others (in fact the same) give the coefficient
$A_m=B_m$. The temperature is finally
\begin{equation}
T(r,z) = \sum_m \frac{\alpha p_m }{K k_m^2} \left(1-\tau
\frac{(\ep^{-k_m(h/2-z)}+\ep^{-k_m(h/2+z)}}
{(\zeta_m+\tau)-(\zeta_m-\tau) \ep^{-k_mh}} \right)\,\, \J_0 (k_m r)
\label{tempsub}
\end{equation}
By using Eq.(\ref{lens}), we find also the thermal lens
\begin{equation}
\Psi(r) = \frac{dn}{dT} \sum_m \frac{\alpha p_m }{K k_m^2} \left( h-
\frac{2\tau}{k_m} \,
\frac {1 - \ep^{-k_m h}}{(\zeta_m+\tau)-(\zeta_m-\tau)\ep^{-k_m h}}
\right) \,\, \J_0 (k_m r).
\label{lenssub}
\end{equation}

\subsection{Numerical examples with VIRGO and VIRGO 'upgrade'}

We give in this part some examples of typical temperature fields
and thermal lensing in the VIRGO context. As the thermal theory is linear,
we first give results separately for the cases of absorption
in the coating and in the substrate for unity absorbed power.
In order to quantify the impact of thermal lensing on the VIRGO performance
we compute, as a figure of merit, the losses ${\cal L}$ induced by the thermal lens.
These losses are computed as the deviation from unity of
the coupling coefficient of a perfect
normalized (power unity) TEM$_{00}$ beam propagating through the mirror, 
so experiencing the thermal lens, to the same TEM$_{00}$ beam. 
The coupling coefficient is nothing but
the squared modulus of the inner product between the two beam complex amplitudes.
We have then
\begin{equation}
{\cal L} = 1 - 
\left|\int_0^\infty \psi_{00}(r) \ep^{ \frac{2i\pi\Psi(r)}{\lambda}} \psi_{00}(r) 2\pi r dr \right|^2
\end{equation}
where $\psi_{00}(r) = \sqrt{2/\pi w_0^2} \exp(-r^2/w_0^2)$ is the amplitude
of the normalized  TEM$_{00}$ beam.

On figure 2 we can see the temperature fields and the corresponding thermal lens
profiles for 1 W absorbed respectively in the coating and in the substrate.
We recall that, although the temperature fields are different for the two cases, with
a maximal temperature rise around 10 K per absorbed Watt in the coating
and less than 3 K per absorbed Watt in the substrate, the thermal lenses are very similar,
with, in both case a maximal optical path difference (OPD) about $3 \mu m$ per absorbed Watt.
In the planned case for VIRGO, we will have
$\epsilon P_{cav} \simeq 10^{-6}\times 15\times 10^3 \simeq 15$ mW absorbed in the
cavity mirror coatings and $\alpha h P_{sub} \simeq 10^{-6} \times 10 \times 10^3
\simeq 10$ mW absorbed in their substrates, so the corresponding maximal OPD is about $7.5\times10^{-2}$
$\mu$m.
If, in case of upgrade, for example the laser power is increased by one order
of magnitude, all the absorbed powers increase accordingly. It will be then rather
100 mW that will be absorbed in the coating and the same amount in the substrate,
so the corresponding maximal OPD is about 0.75
$\mu$m.
If now we compute the induced losses in the case of VIRGO, we find
${\cal L} \simeq 1.7\times 10^{-3}$, which is reasonable, and in the case
of the upgrade ${\cal L} \simeq 0.16$ which is {\bf disastrous}.
The situation seems a little better if we discard the parabolic part (obtained
by a fit of the thermal lens in the mirror central part) from the thermal lens.
Indeed, a parabolic lens effect can be compensated for by a proper matching
lens at the input of the interferometer. Of course, this works only if the
thermal lensing is well symmetric between the input mirrors of the kilometric
cavities, otherwise it will be impossible to compensate for parabolic
parts of both thermal lenses by some input matching optics.
In this case, the losses
become ${\cal L} \simeq 2.5\times 10^{-4}$ in the case of VIRGO
and, in the case of VIRGO upgrade, ${\cal L} \simeq 2.2\times 10^{-2}$ 
which is still too high.
These results clearly call for decreasing the thermal gradients in case of
(even minimal) upgrade.

NB: in the following, all the numerical results involves the ``VIRGO upgrade''
case ($P_{sub} \simeq 10$ kW and $P_{coat} \simeq 150$ kW) and the up-to-date
values for the absorption coefficients ($\epsilon \simeq 1$ ppm and
$\alpha\simeq 1$ ppm/cm), except the final ones given for the ``advanced LIGO''
configuration.

\section{Compensation of thermal effects by absorption in the 
substrate of a ring-shaped beam}

In this section, we will try to solve the thermal gradient problem by heating
the external part of the mirror by a ring-shaped beam, in addition of the 
main Gaussian beam, which is always supposed to be absorbed in the coating
and in the substrate.
We model the intensity of the extra beam as
\begin{equation}
I_e(r) = \begin{cases} 0 & \text{if} \,\,\, r <  a_e \\
I_0 & \text{if} \,\,\, a_e \leq r \leq a
\end{cases}
\end{equation}
where $I_0$ is constant. The total power
in the beam is then
\begin{equation}
P_e = I_0 \pi (a^2-a_e^2)
\end{equation}
If the ring-shaped heating zone is obtained with
the help of a a number of diodes located all around the mirror, then the
quantity $a-a_e$ represents nothing but the penetration length (related
to the lineic absorption of the diode wavelength) of these
beams in the substrate.

There is thus an extra source of heating in the substrate with respect to 
the previous case. The temperature field due to absorption of light
in the substrate is then the superposition of the one computed in section 2
and the one due to absorption of the ring-shaped beam. For the later, the solution
is essentially the same, except that $\alpha$ has to be replaced by the lineic
absorption $\alpha_e$ for the wavelength of the extra beam, and the coefficients
of the Dini expansion of the intensity have to be changed.
Let's note  $p^{(e)}_m$ the coefficients related to $I_e(r)$.
According to Eq.(\ref{pm}), we have then
\begin{equation}
p^{(e)}_m = \frac{2}{a^2}\frac{1}{1+\tau^2/\zeta_m^2}\frac{I_0}{\J_0 (\zeta_m)^2}
\int_{a_e}^a  \J_0 (k_m r) r dr.
\end{equation}
By direct integration of $x\J_0 (x)$ in $x \J_1 (x)$ and by substitution of 
$I_0$ by the corresponding  power $P_e$, we have finally
\begin{equation}
p^{(e)}_m = \frac{P_e}{\pi(a^2-a_e^2)}\frac{2\zeta_m}{\tau^2+\zeta_m^2}
\frac{1}{\J_0 (\zeta_m)^2} \left( \J_1 (\zeta_m) - \frac{a_e}{a} \J_1(\zeta_m a_e/a)\right)
\end{equation}
From Eq.(\ref{tempsub}), we can then write the total temperature field due to absorption
in the substrate:
\begin{equation}
T(r,z) = \sum_m \frac{\alpha p_m +\alpha_e p^{(e)}_m}{K k_m^2} \left(1-\tau
\frac{(\ep^{-k_m(h/2-z)}+\ep^{-k_m(h/2+z)}}
{(\zeta_m+\tau)-(\zeta_m-\tau) \ep^{-k_mh}} \right)\,\, \J_0 (k_m r)
\end{equation}
In the same way, we obtain the thermal lensing profile by substituting $\alpha p_m$
by $\alpha p_m + \alpha_e p^{(e)}_m$ in Eq.(\ref{lenssub}).
The expressions for the temperature field and thermal lens
due to absorption in the coating are of course unchanged.

The impact of the supplementary heating by the extra beam depends on two parameters,
the radius $a_e$ which indicates which size of the mirror is heated and the power
of the beam, or, better, the absorbed power in the mirror $\alpha_e h P_e$.
In order to optimize the situation, figure 3 shows the mirror losses as a function
of the radius $a_e$, the absorbed power being fixed for each value of $a_e$ to the
value that minimizes thes losses. We see clearly that there is an optimum at
$a_e \simeq 2$ cm, that is the numerical value of the waist $w_0$ of the main beam.
Henceforth, we set then $a_e = 2$ cm. The figure 4 shows the mirror losses as a function
of the absorbed power. We see then that the heating by the ring-shaped beam
dramatically improves the situation: the losses can be decreased from the (inacceptable)
value ${\cal L} \simeq 0.16$ down to ${\cal L} \simeq 2.7\times 10^{-3}$, so a 'gain'
of about a factor 60 for the losses. An extra factor about 2 (corresponding
to losses as low as $1.2\times 10^{-3}$) can be gained if we substract the parabolic part
of the thermal lens. Heating the cold parts of the mirror seems then of great benefit, but
the obvious drawback is that a large amount of light power has to be dissipated, about
38 W for the optimal case (that gives a dissipated intensity
$I_o \simeq$ 3 W/cm$^2$), that means a source of yet higher power.
If we imagine a battery of laser diodes around the mirror, delivering say 1 W each, 
at least 38 of them will be needed.
In order to understand why so large powers are needed, let's take a look at the
shapes at the profiles of the thermal lenses in the optimal
case and for too low or too high absorbed powers (see fig.5)
 We note that the total thermal lens has a quite perfect flat profile at the center
of the mirror. This compensation for the thermal lensing due to absorption of the main beam
can be only achieved for a particular power absorbed from the extra beam. If the absorbed
power is too low or too high, then there remains a concave or convex lens.

\section{Compensation by absorption in the 
substrate of an attenuated ring-shaped beam}

We now consider the physically different case of an external ring-shaped
beam propagating parallely to the main beam and being (strongly) absorbed by
the mirror substrate. We just saw in the previous section that, in order to be efficient,
the external beam must deposit a large amount of power in the substrate. That indicates
that in the case now considered the extra beam will be strongly attenuated along its
propagation through the mirror.
We have then to solve the heat equation with an extra source of heat being the
absorbed light intensity. As in the previous section we adopt a ring-shaped profile
for the intensity but with now an attenuation along z.
We suppose that the new light source is in the intra-cavity side, but it can be in the other
side, it will give exactly the same new thermal lens.
The intensity can be then written as
\begin{equation}
I_e(r,z) = \begin{cases} 0 & \text{if} \,\,\, r <  a_1 \\
I_0 \ep^{-\alpha_2(z+h/2)}& \text{if} \,\,\, a_1 \leq r \leq a_2
\end{cases}
\end{equation}
where $\alpha_2$ is the lineic absorption of Silica at the wavelength of the new light beam.
Note that by chosing wavelengths near 0.5, 1.4 or 3 $\mu$m (OH absorption bands),
we can have values of $\alpha_2$ in the range 0.1 to 0.95 /cm \cite{cor,lorr}
We note also that we can have a priori an external radius $a_2$ of the beam lesser than
the mirror radius. 

We have now to solve the Heat equation:
\begin{equation}
-K \Delta T = \alpha_2 I_e(r,z)
\label{heatatt}
\end{equation}
The general solution of the homogeneous equation has always the form of Eq.(\ref{sol}).
We have to add a particular solution of (\ref{heatatt}) to the later.
Such as a particular solution can be looked for 
under the form $\sum t_m \ep^{-\alpha_2z} \J_0 (k_m r)$ and we find:
\begin{equation}
t_m = \frac{\alpha_2}{K} \frac{p^{(e)}_m}{k_m^2-\alpha_2^2} \ep^{-\alpha_2 h /2}.
\end{equation}
The solution is finally:
\begin{equation}
T(r,z) = \sum_m \left(A_m \ep^{k_m z}+B_m \ep^{-k_m z}+ \frac{\alpha_2}{K} 
\frac{p^{(e)}_m}{k_m^2-\alpha_2^2} \ep^{-\alpha_2 (h /2+z)}\right) \J_0 (k_m r).
\end{equation}
The boundary condition Eq.(\ref{cl1}) gives again that $\zeta_m = k_m \times a$
is the m-th zero of $x\J_1 (x) - \tau \J_0 (x)$, and the two others Eq.(\ref{cl2}) (with
$\epsilon = 0$)
and Eq.(\ref{cl3}) give:
\begin{align}
A_m &= \frac{\alpha_2}{K} \frac{p^{(e)}_m}{k_m^2-\alpha_2^2}\,\, \ep^{-\alpha_2 h /2} \,\,
\frac{(\alpha_2a-\tau)(\tau+\zeta_m)\ep^{-\alpha_2h}+(\alpha_2a+\tau)(\tau-\zeta_m)\ep^{-k_mh}}
{(\zeta_m+\tau)^2-(\zeta_m-\tau)^2 \ep^{-2k_mh}} \\
B_m &= - \frac{\alpha_2}{K} \frac{p^{(e)}_m}{k_m^2-\alpha_2^2}\,\, \ep^{-\alpha_2 h /2} \,\,
\frac{(\alpha_2a+\tau)(\tau+\zeta_m)+(\alpha_2a-\tau)(\tau-\zeta_m)\ep^{-(\alpha2+k_m)h}}
{(\zeta_m+\tau)^2-(\zeta_m-\tau)^2 \ep^{-2k_mh}}
\end{align}
The thermal lens due this temperature field is derived from Eq.(\ref{lens}):
\begin{equation}
\Psi(r) = \frac{dn}{dT} \sum_m \frac{p^{(e)}_m}{K(k_m^2-\alpha_2^2)}
\left(\frac{\alpha_2}{k_m}  
\frac{(\alpha_2a-\tau)\ep^{-\alpha_2h}-(\alpha_2a+\tau)}{(\zeta_m+\tau)-(\zeta_m-\tau)\ep^{-k_mh}}
\left(1-\ep^{-k_mh}\right)+\left(1-\ep^{-\alpha_2h}\right)\right)
\J_0 (k_m r).
\end{equation}
We obtain then the total temperature field by superposition of the field due to absorption
of the main beam (in the coating and in the substrate) and of the field due to absorption
of the auxiliary ring-shaped beam. The same for the thermal lens.
Let's notice that the absorbed power in the substrate  from the auxiliary beam is
\begin{equation}
P_{abs} = \int_{-h/2}^{h/2} \alpha_2 \ep^{-\alpha_2(h/2+z)} P_0 dz = P_0 \left(1-\ep^{-\alpha_2h}\right)
= \alpha_2 h P_0 \,\, \left( \frac{1-\ep^{-\alpha_2h}}{\alpha_2h}\right)
\end{equation}
where $P_0=2\pi \int I_0\, r dr = I_0 \pi (a_2^2-a_1^2)$.

There are now four free parameters in this part : the extra beam inner radius $a_1$ and outer radius
$a_2$, the power of the source $P_0$ and the absorption $\alpha_2$ at the wavelength of the source.
The two last ones can be merged into a single one, that is the absorbed power. We will see indeed that
the results actually do not depend on $P_0$ and $\alpha_2$ but simply on $P_{abs}$.

As a first example, figure 6 shows the losses induced by the total thermal lens as a function of $P_{abs}$
for $a_1 =2$ cm, $a_2=a$ and $\alpha_2 =0.2$ /cm, that is a penetration 
factor $ 1 - \exp(-\alpha_2 h) \simeq 0.865$. We note that the minimal losses are (again) of the order
of $2.7\times10^{-3}$ and are obtained with $P_{abs} \simeq 38$ W.
If we allow now $\alpha_2$ to vary, we find in fact the same kind of results : minimal losses about
$2.7\times10^{-3}$ for the same absorbed power about 38 W. If we optimize now with respect to $a_1$ and $a_2$, we find, just
as in the previous section, an optimum at $a_1 \simeq 2$ cm (corresponding to the size $w_0$ 
of the main beam) and at $a_2 \simeq 19$ cm, just below the mirror radius $a \simeq 20$ cm.

The minimal losses achievable by this method are ${\cal L} \simeq 2.6\times 10 ^{-3}$, and are given
by $a_1\simeq 2$ cm, $a_2\simeq 19$ cm and whatever the value of $\alpha_2$. 
Of course the higher $\alpha_2$, the
smaller the needed source power $P_0$. It has to be noticed that the parabolic part removal from the
thermal lensing doesn't improve the minimal losses here. 
There is actually a gain when the absorbed power is well below
or above, but not around the optimal value (about 38 W), for which the lens is almost perfectly flat at the center
of the mirror and parabolic removal can not improve anything anymore. The minimal losses quoted above
are then an absolute limit for this method.

The conclusions of this section are then similar to the previous one, except that there can not be
any extra gain due to parabolic lens removal. In both case, in the optimal absorption case, the
mean temperature of the mirror is about 20 K. The approximation of low heating ($T\ll T_e$) is
still reasonable (error on the radiative flux calculation less than 10\%).

\section{Compensation by fixing the mirror circumference temperature}

\subsection{Heat Equation general solution and boundary conditions}
We suppose in this section that the temperature of the mirror circumference is fixed to a
temperature $T_0$ by some thermostat (for instance an electric blanket or some equivalent
device). We always suppose that the vacuum tank is at the temperature $T_e$, that is
the temperature 'seen' by the mirror sides at $z=\pm h/2$. From now on,
we will take $T_0$ as the reference temperature, and $T(r,z)$ will then denote
the deviation from this reference.
The heat flux,Eq.(\ref{flux}) is in this case:
\begin{equation}
F=\sigma \left((T_0+T)^4-T_e^4   \right) \simeq 4\sigma T_e^3 T + 4\sigma T_e^3(T_0-T_e)
\end{equation}
We will then have the same expression for the radiative flux on the mirror sides, except
that appears a supplementary (constant) term that we will note $C = 4\sigma T_e^3(T_0-T_e)$.
The boundary conditionds can now be expressed as:
\begin{align}
- &K \frac{\partial T}{\partial r}(r=a, z) = 0 \label{cl1b}\\
- &K \frac{\partial T}{\partial z}(r, z = h/2) = 4e\sigma{T_e}^3 T(r, z=h/2)+C\label{cl2b}\\
- &K \frac{\partial T}{\partial z}(r, z =-h/2) = \epsilon I(r) - 4e\sigma{T_e}^3 T(r, z=-h/2)-C\label{cl3b}
\end{align}
We solve the Heat Equation as in section 2. A general solution of the homogeneous Heat
Equation is always given by Eq.(\ref{sol}). The first boundary condition Eq.(\ref{cl1b}) implies
that the $k_m$ are now defined as $\zeta_m=k_m\times a$ are the zeros of $\J_0 (x)$.
The functions $\J_0 (k_m r)$ form again a complete orthogonal set with the normalization
condition given by \cite{handbook}
\begin{equation}
\int_0^a \J_0 (k_m r) \J_0 (k_n r) r dr = \frac{1}{2} a^2  \J_1 (\zeta_m)^2  \delta_{nm}. \label{normb}
\end{equation}
The Gaussian beam intensity can always be expanded on this set, 
and the expression of the $p_m$ has to be
modified in accord the new normalization constants :
\begin{equation}
p_m = \frac{P}{\pi a^2} \frac{1}{\J_1 (\zeta_m)^2} \exp\left( \frac{-k_m^2 w_0^2}{8}\right)
\end{equation}
with the same notations as in section 2.

In the two other boundary conditions, the constant $C$ can be expanded on the set of the
$\J_0 (k_m r)$. If we note $C= \sum c_m \J_0 (k_m r)$, we have:
\begin{equation}
c_m = \frac{2}{ a^2 \J_1 (\zeta_m)^2} \int_0^a C \J_0 (k_m r) r dr.
\end{equation}
We note that this amounts to compute the Hankel transform of a radial window function, 
and after a simple calculation, using that the derivative of $x\J_1 (x)$ is $x\J_0 (x)$, we find
\begin{equation}
c_m = \frac{2 C}{ \zeta_m \J_1 (\zeta_m)} 
\end{equation}
In practice, the number $N$ of terms to consider in the Dini Series will be fixed by the condition
\begin{equation}
1 \simeq \sum_{m=1}^N \frac{2 \J_0 (k_m r)}{ \zeta_m \J_1 (\zeta_m)} \,\,\, \forall r \in [0,a[.
\end{equation}
From figure 7, we see there is an optimum at $N\simeq 700$, and, consequently we will chose
$N=700$ in the following. Note that for all the previous numerical results, we have chosen
$N=50$, which was actually comfortable.

\subsection{Absorption in the coating}

In the case of absorption in the coating, the boundary condtions Eq.(\ref{cl2b}) and
Eq.(\ref{cl3b}) become:
\begin{align}
 -K\left( k_m A_m \ep^{k_m h /2} - k_m B_m \ep^{-k_m h/2}\right) & = 
\frac{\tau K}{a}\left(A_m \ep^{k_m h /2}+B_m \ep^{-k_m h/2}\right) +c_m \\
-K\left( k_m A_m \ep^{-k_m h /2} - k_m B_m \ep^{k_m h/2}\right) & = 
\epsilon p_m -\frac{\tau K}{a}\left(A_m \ep^{-k_m h /2}+B_m \ep^{k_m h/2}\right) -c_m
\end{align}
The solution of the system is then, after a straightforward calculation,
\begin{align} 
A_m &= \frac{a}{K} \ep^{-k_mh/2} \frac{(\epsilon p_m- c_m)(\zeta_m -\tau)\ep^{-k_mh}-c_m(\zeta_m +\tau)}
{(\zeta_m+\tau)^2-(\zeta_m-\tau)^2 \ep^{-2k_mh}} \\
B_m &= \frac{a}{K} \ep^{-k_mh/2} \frac{(\epsilon p_m- c_m)(\zeta_m+\tau)-c_m(\zeta_m-\tau)\ep^{-k_mh}}
{(\zeta_m+\tau)^2-(\zeta_m-\tau)^2 \ep^{-2k_mh}}
\end{align}
So we have the solution of the Heat Equation for heating in the coating by absorption of the gaussian beam
by the Dini series Eq.(\ref{sol}).

The thermal lens is again calculated from Eq.(\ref{lens}) and from the temperature Dini expansion:
\begin{equation}
\Psi(r) = \frac{dn}{dT}\frac{a}{K} \sum_m \frac{\epsilon p_m -2c_m}{k_m}
\frac {1 - \ep^{-k_m h}}{(\zeta_m+\tau)-(\zeta_m-\tau)\ep^{-k_m h}} \, \J_0 (k_m r).
\end{equation}
Of course, if $C=0$, that is all the $c_m$ vanish, we find again the expression of Eq.(\ref{coatlens}),
the only difference being then the definition of the $k_m$.

\subsection{Absorption in the substrate}

The solution of the Heat Equation Eq.(\ref{heat}) is again the superposition of the general solution
of the homogeneous equation $\Delta T = 0$ and a particular solution.
The solution has again the form of Eq.(\ref{subsol}), and, this time,
 the two boundary conditions Eq.(\ref{cl2b})
and Eq.(\ref{cl3b}) give:
\begin{equation}
A_m=B_m=  \frac{ \frac{\tau\alpha p_m}{K k_m^2} + \frac{a c_m}{K}} 
{(\zeta_m+\tau)-(\zeta_m-\tau) \ep^{-k_mh}} \,\, \ep^{-k_mh/2}
\end{equation}
so that the temperature is:
\begin{equation}
T(r,z)= \sum_m  \left(\frac{\alpha p_m}{K k_m^2}-\frac{ \frac{\tau\alpha p_m}{K k_m^2} + \frac{a c_m}{K}} 
{(\zeta_m+\tau)-(\zeta_m-\tau) \ep^{-k_mh}} \left( \ep^{-k_m(h/2+z)}+\ep^{-k_m(h/2-z)}\right) \right) \J_0 (k_mr)
\end{equation}
After integration of Eq.(\ref{lens}), we find the thermal lens profile:
\begin{equation}
\Psi(r) = \frac{dn}{dT}\sum_m  \left(\frac{\alpha h p_m}{K k_m^2}-\frac{2}{k_m}
\frac{ \frac{\tau\alpha p_m}{K k_m^2} + \frac{a c_m}{K}} {(\zeta_m+\tau)-(\zeta_m-\tau) \ep^{-k_mh}} 
\left( 1-\ep^{-k_mh}\right) \right)\J_0 (k_m r).
\end{equation}
If all the $c_m=0$, we find again the same expression as Eq.(\ref{lenssub}).

\subsection{Numerical results}

The only free parameter in this compensation method is the temperature $T_0$ of the thermostat (if the
temperature $T_e$ of the surrounding vacuum tank - the tower - is fixed).
The figure 8 shows the total losses as a function of the thermostat temperature. We note there is a minimum
occuring for $T_0 \simeq 311$ K, so a temperature rise about 15 K with respect to the tower temperature.
But the minimum is rather high, about ${\cal L} \simeq 1.2\times 10^{-2}$, so one order of magnitude
larger than the minimal losses obtained with the other compensation methods described in previous sections.
Clearly this method is not competitive. Removing the parabolic part does not help much here and changing the
temperature $T_e$ does not change roughly the results. For sake of completeness, figure 9 shows the temperature
and thermal lens profile for the optimal thermostat temperature ($T_0\simeq 311$ K). We note an average temperature
mirror less than 10 K and a roughly (but not sufficiently) flat profile of the thermal lens at
the center of the mirror

\section{Sapphire substrates}

It is already well known that replacing Silica substrates by Sapphire ones can already improve the situation.
For example, for the VIRGO 'upgrade' situation, the losses due to the thermal lensing can be decreased
from ${\cal L} \simeq 0.16$ to ${\cal L} \simeq 1.2\times 10^{-3}$ if we simply replace Silica by Sapphire.
This is simply due to the much larger conductivity of Sapphire ($K\simeq 33.0$ W.m$^{-1}$.K$^{-1}$ vs
$K\simeq 1.38$ W.m$^{-1}$.K$^{-1}$ for Silica).
If we use in addition a compensation method, we can have even better results. In the case of compensation
by absorption of a ring-shaped beam (sections 3 and 4), the losses can be decreased 
down to ${\cal L} \simeq 1.7 \times 10^{-5}$, that is 17 ppm ($1.1 \times 10^{-5}$ - 11 ppm - with parabolic part removal). The drawback
is that this optimum is reached with a larger absorbed power $P_{abs} \simeq 46$ W than in the Silica
case ($P_{abs} \simeq 38$), but it is not a dramatical increase of the needed extra-source light power.
Note that in the case of compensation with the help of a thermostat (section 5), we can achieve losses
less than $10^{-3}$, but this requires a fine tuning of the thermostat temperature by
a fraction of degree in excess from the tower temperature, something non realistic (for example if compared
to the tower temperature inhomogeneities).

\section{Influence of the beam size}

Increasing the beam size is also a way to decrease the intensity absorbed in the mirrors, and so a way
to decrease the thermal gradients. In this section, we are going to study the influence of the (main) beam
size on the thermal losses and, in particular, the impact of a compensation method. Note that in the ``advanced
LIGO'' concept, large beam sizes are planned to be used, typically $w_0 \simeq 5.5$ cm in case of Silica substrates \cite{ligo2bis}.
We show only results for the compensation method by absorption of a ring-shaped beam (see sections 3 and 4), since the last
method (section 5) is not competitive. 
For each beam size, the minimal losses are found (1) by optimizing the inner radius of the ring-shaped
beam (2) by optimizing the absorbed power. We find that the minimal losses are roughly independant of the main beam size, but, what
is changing is the optimized absorbed power from the extra beam. The minimal losses are ${\cal L} \simeq 2\times 10^{-3}$ 
(if $w_0 \geq 3$ cm) and the
needed absorbed power decreases down to about 4.5 W for a beam size $w_0=6$ cm, instead of 38 W for a beam size $w_0=2$ cm, as
shown by figure 10.
Increasing the beam size then does not change the minimal possible losses, but it changes the needed power to reach them.
Note that the optimized inner radius of the ring-shaped beam is generally found to be of the order of the main beam size $w_0$
(so that the cold parts of the mirror are effectively heated by the compensation beam).

\section{The case of the advanced LIGO configuration}

It is interesting to state the potential of compensation methods with a 2nd generation configuration for which the
design is well advanced and precise numbers are available.
For example, the ``advanced LIGO'' design plans either Silica substrates with  absorption
about 0.5 ppm/cm, mirror dimensions $a\simeq 19.4$ cm and $h\simeq 15.4$ cm, a beam size $w_0\simeq 5.5$ cm, 
1.4 kW crossing the substrates and 530 kW stored in the cavities or Sapphire substrates
with absorption around 40 ppm/cm, mirror dimensions $a\simeq 15.7$ cm and $h\simeq 13$ cm, 
a beam size $w_0\simeq 6$ cm, 2.1 kW crossing the substrates and 830 kW stored in the cavities.
For both cases, the coating absorption is expected at the level of $\epsilon \simeq 0.5$ ppm, hence a factor 2 better
than today's best coatings.
With such numbers, and without thermal compensation, the losses for Silica substrates are about 0.15, while they are
about $1.8\times 10^{2}$ for Sapphire substrates. It is clear again that thermal compensation is needed for this
precise configuration, whatever the substrate materials, unless absorptions can be drastically lowered. It is not likely
to be much the case for Silica, but prospects for Sapphire are more optimistic, and absorption around 20 ppm/cm or less, as already
observed in small samples \cite{bl}, is not out of reach.
With thermal compensation using an extra ring-shaped beam, the minimal losses achievable with Silica substrates are found to be
about $5\times 10^{-3}$ (less than $2\times 10^{-3}$ with removal of the parobolic part). This result means already
(at least) losses around $1 \%$  only due to thermal lensing in input mirrors of the kilometric cavities. This is not encouraging
for using Silica substrates.
With Sapphire substrates now, the figure 12 shows the minimal losses as a function of the absorption in the range 10-50 ppm/cm
(since it is today like a free parameter that will probably evolve favourably). Already with absorption about 40 ppm/cm,
Sapphire does better than Silica, with losses about $10^{-3}$ ($2\times 10^{-4}$ with parabolic part removal). With absorption
about 20 ppm/cm, the losses decrease down to about 460 ppm (some tens of ppm with parabolic part removal). With Sapphire substrates
and thermal compensation, thermal lensing should not be longer a source of concern for the good working of the interferometer.

\section{Conclusions}

In this paper, we have theoretically investigated several means to compensate for thermal effects in 
mirrors of (advanced) interferometric detectors of gravitational waves. 
The best method seems to heat the cold parts of the substrates by absorption of a ring-shaped
beam, while the method of thermostating the circumference of the mirror is clearly not competitive.
 With the VIRGO numbers, Silica substrates and absorption coefficients $\epsilon=1$ ppm and
$\alpha=1$ ppm/cm, and in case of increase of the circulating powers by a factor 10, the losses can
be lower down to ${\cal L} \simeq 2.7\times 10^{-3}$, vs ${\cal L} \simeq 0.16$ if nothing is done.

A factor about 2 can still be gained if we can remove the parabolic part of the thermal lens, but this supposes
that the thermal lenses are well symmetric between the two arms of the interferometer.
For a beam size $w_0 =2$ cm, the order of magnitude of the light source power needed to achieve the minimal losses value 
is larger than about 40 W, since an {\it absorbed} power of about 38 W is required. This needed power can be lowered
with a larger beam size. For instance, only 4.5 W are needed to be absorbed if $w_0=6$ cm, that means a gain of almost
one order of magnitude for the required power. Of course the beam size cannot be increased much further due to the finished
size of the mirrors (typically not larger than 20 cm).

Now, again with the example of the VIRGO numbers,
if we want a gain of one order of magnitude for the shot-noise limited sensitivity, that means an increase
of the circulating powers by a factor 100 this time. If the absorption coefficients in the coatings and in the
Silica substrates can not be decreased by at least one order of magnitude, it is then clear that we will 
have to give up the Silica for the substrates, in favour of  Sapphire for instance, since it will no longer
be possible to compensate for thermal gradients (the compensation methods can not do miracles: eg with
$P_{coat} \sim 1$ MW, $P_{sub} \sim 100$ kW, $\epsilon=1$ ppm
and $\alpha=1$ ppm/cm, the losses can not be lower than 97\% !). 
This means finally that the mirror R\&D effort needed for future
upgrades has two options: either going on with Silica susbtrates and improve absorption 
coefficients at 1.064 $\mu$m (better coating technology, lesser OH radicals in the substrate ...)
or stopping with Silica technology and developping substrates with other materials like Sapphire.
Numerical results obtained with the advanced LIGO configuration are unambiguous regarding the substrate materials :
Sapphire is (will) be better. 

  

\newpage
\begin{figure}
\centerline{\epsfig{file=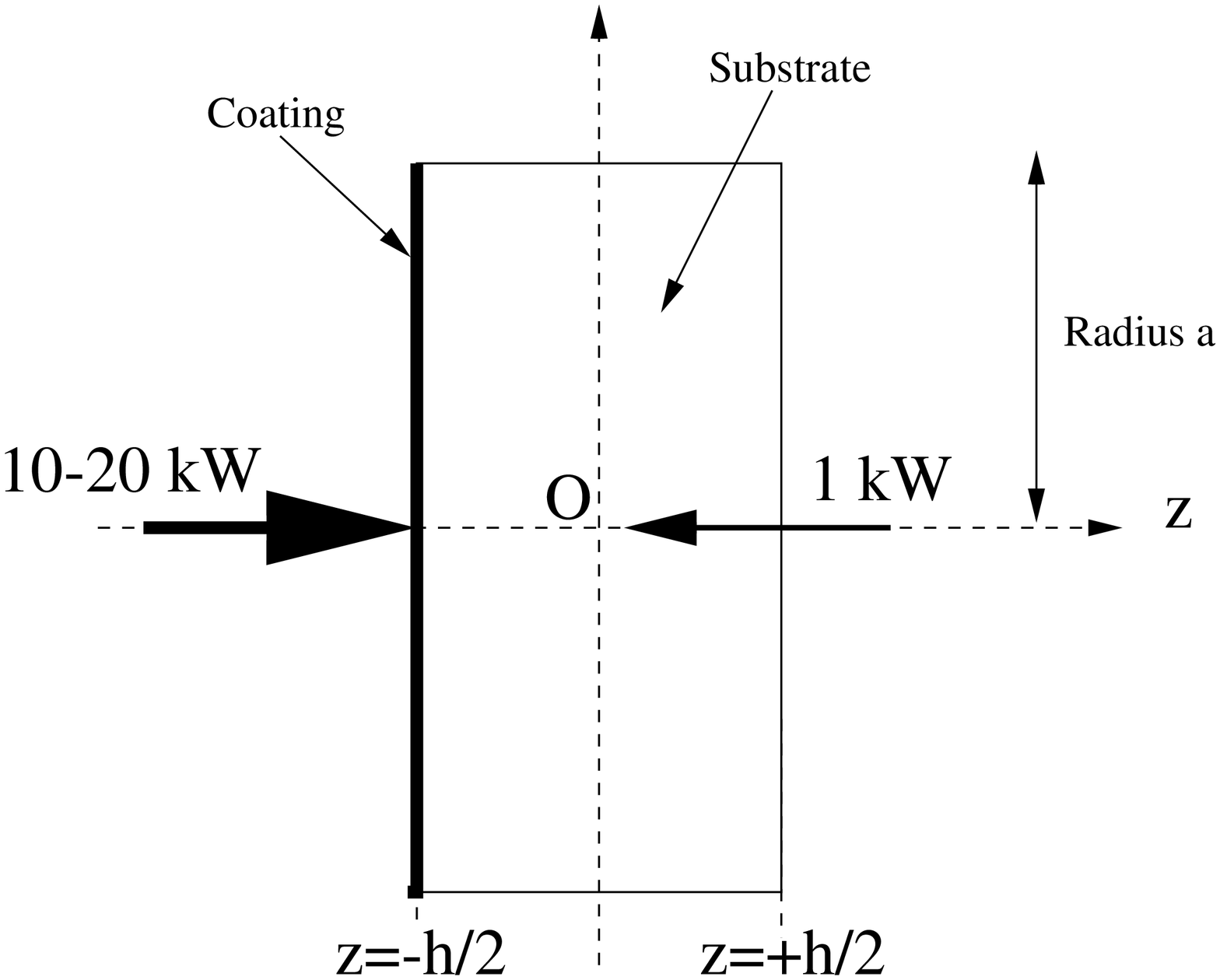,height=10cm}} 
\caption{Heating of a cylindrical mirror by a Gaussian beam. The mirror can
absorb light power in its coating (cavity side) and in its substrate.}
\label{f1}
\end{figure}

\begin{figure}
\centerline{\epsfig{file=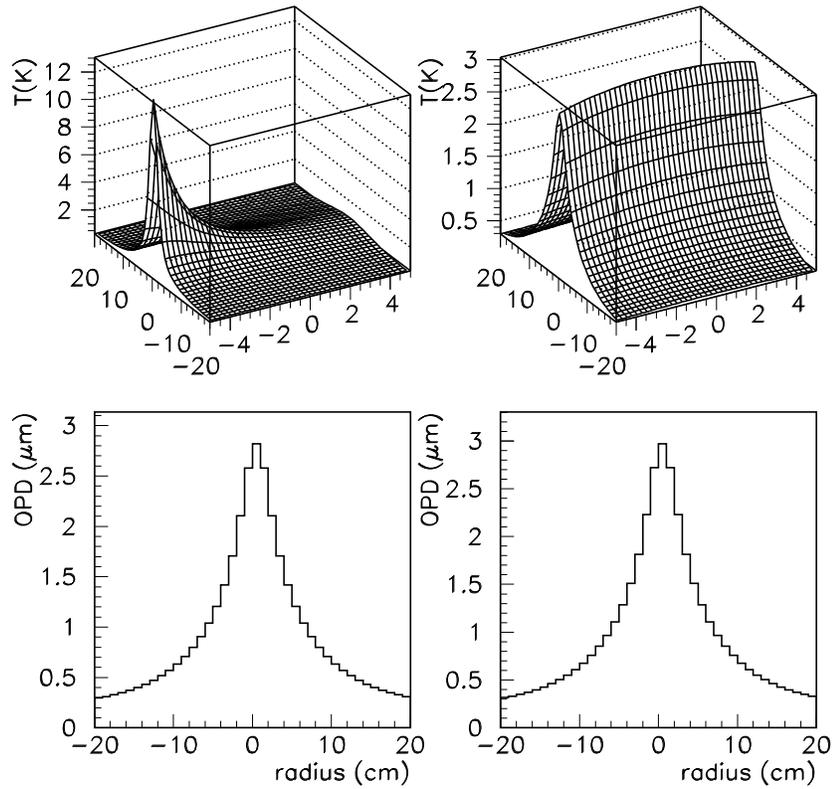,height=12cm}} 
\caption{Upper left: temperature field for 1 W absorbed in the coating.
Upper right: temperature field for 1 W absorbed in the substrate.
Lower left: the thermal lens profile (optical path difference) 
for 1 W absorbed in the coating. 
Lower right: the thermal lens profile for 1 W absorbed in the substrate.}
\end{figure}

\begin{figure}[H]
\centerline{\epsfig{file=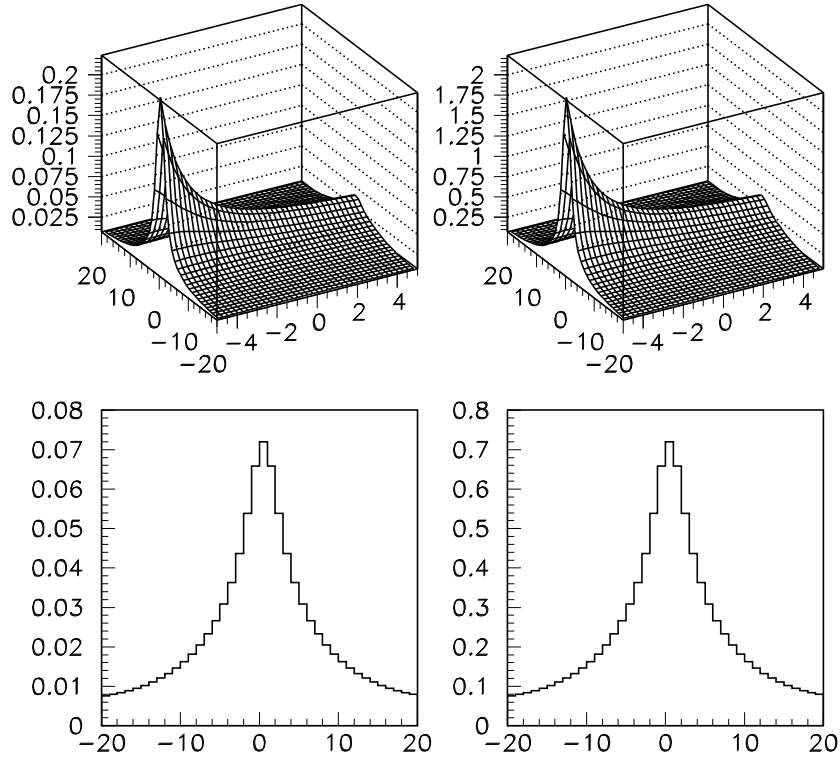,height=12cm}} 
\caption{Upper left: temperature field for 15 mW absorbed in the coating
and 10 mW in the substrate (typical VIRGO numbers). The temperature scale is in K.
Upper right: temperature field for 150 mW absorbed in the coating and  100 mW
in the substrate (VIRGO upgrade numbers). The temperature scale is in K.
Lower left: the thermal lens profile (optical path difference) 
for 15 mW absorbed in the coating
and 10 mW in the substrate (typical VIRGO numbers). The OPD scale is in $\mu m$. 
Lower right: the thermal lens profile for 150 mW absorbed in the coating and  100 mW
in the substrate (VIRGO upgrade numbers). The OPD scale is in $\mu m$.}
\end{figure}

\begin{figure}[H]
\centerline{\epsfig{file=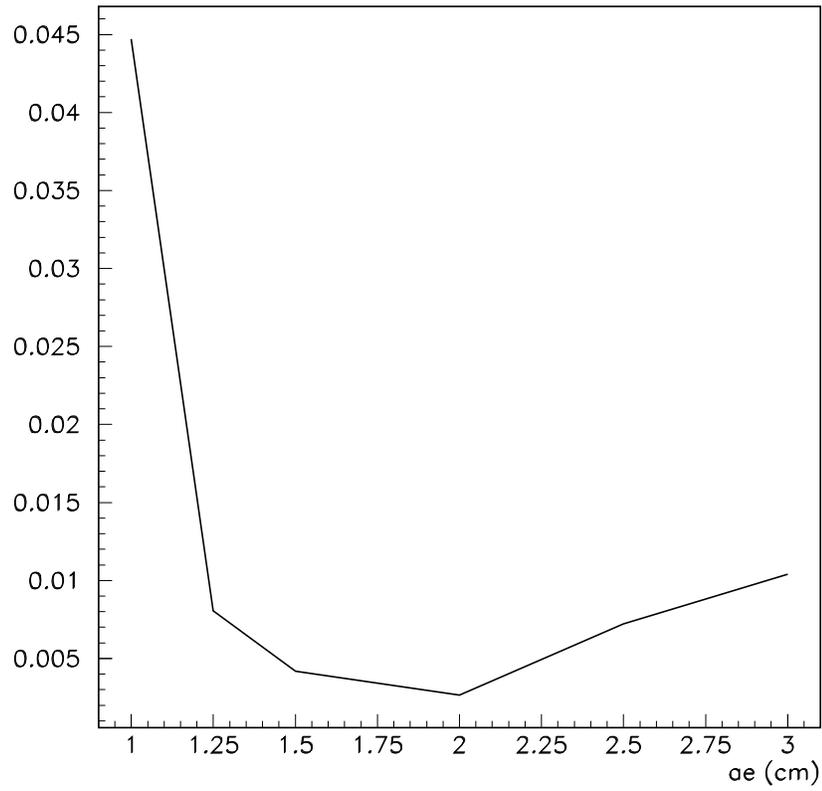,height=12cm}} 
\caption{Losses vs the inner radius $a_e$ of the auxiliary beam.}
\end{figure}

\begin{figure}[H]
\centerline{\epsfig{file=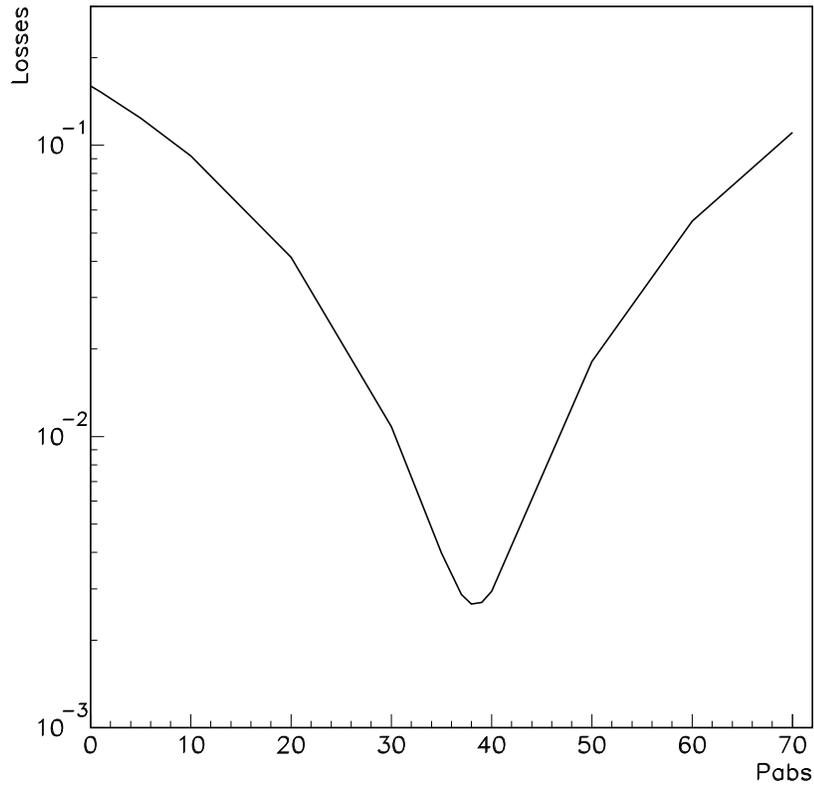,height=12cm}} 
\caption{Losses a as a function of the power absorbed from the extra beam. The losses
are minimal for the quite large power $P_{abs} \simeq 38$ W.}
\end{figure}

\begin{figure}[H]
\centerline{\epsfig{file=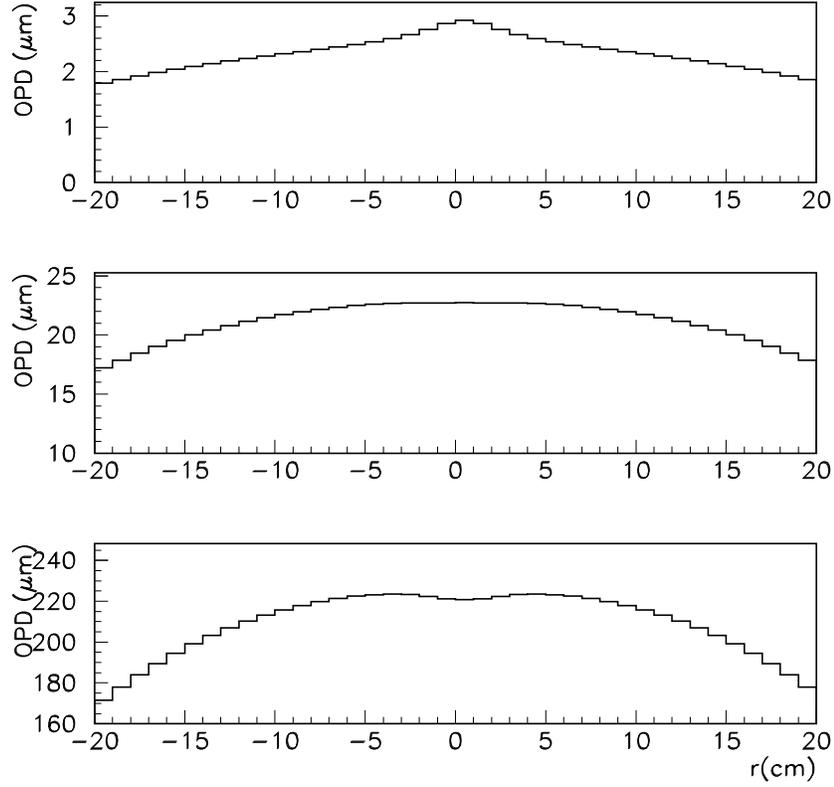,height=12cm}} 
\caption{Thermal lens profiles for respectively $P_{abs} \simeq 3.8 W$ (10 times too low),
$P_{abs} \simeq 38 W$ (optimal case) and $P_{abs} \simeq 380 W$ (10 times too high).
In the optimal case, we note the flatness of the lens in the central part of the mirror.}
\end{figure}

\begin{figure}[H]
\centerline{\epsfig{file=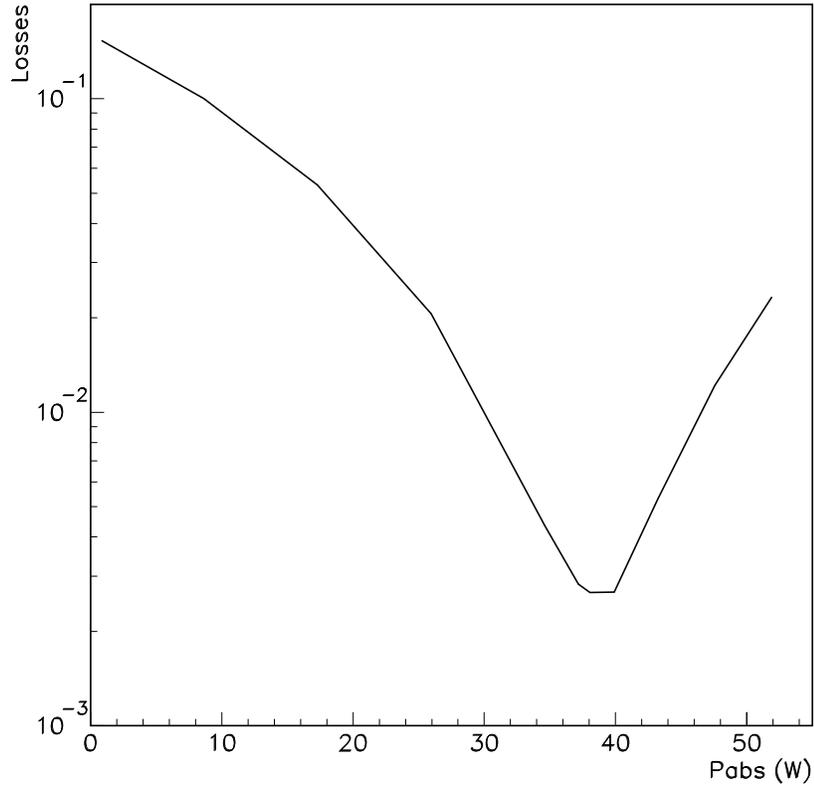,height=12cm}} 
\caption{Losses a as a function of the power absorbed from the extra beam. The parameters are
$a_1=2$ cm, $a_2= 20$ cm  and $\alpha_2=0.2$/cm. The losses
are minimal for the quite large power $P_{abs} \simeq 38$ W.}
\end{figure}

\begin{figure}[H]
\centerline{\epsfig{file=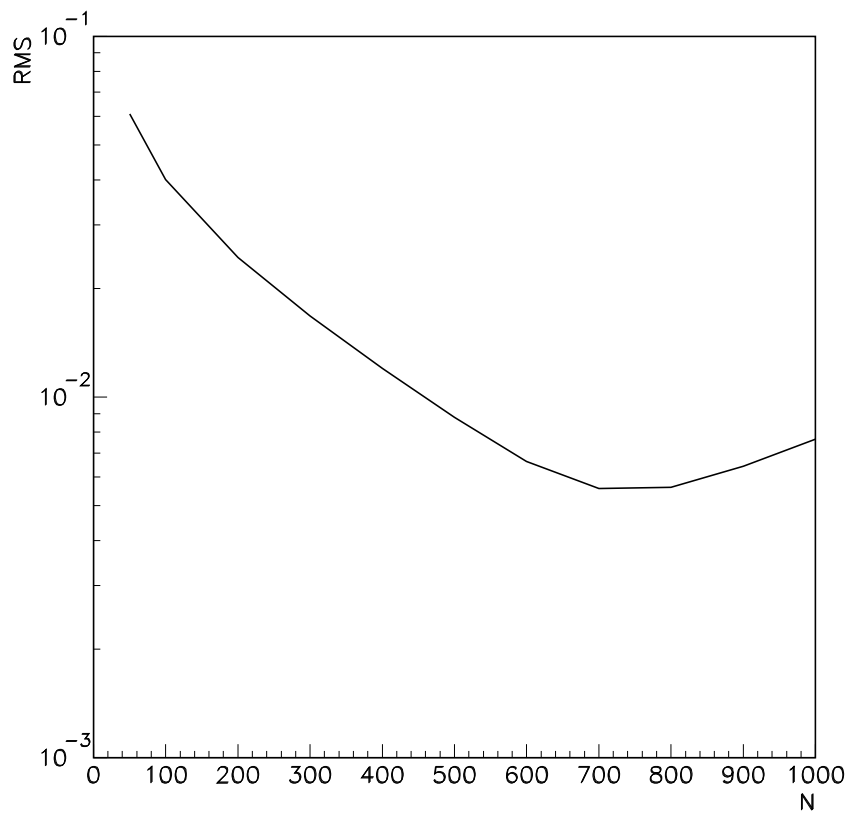,height=12cm}} 
\caption{RMS of the distribution of $\sum 2 \J_0 (k_m r)/  \zeta_m \J_1 (\zeta_m)$ for a sampling
of $r$ in $[0,a]$ as a function of $N$, number of terms in the series. The RMS is minimal
for $N\simeq 700$.}
\end{figure}

\begin{figure}[H]
\centerline{\epsfig{file=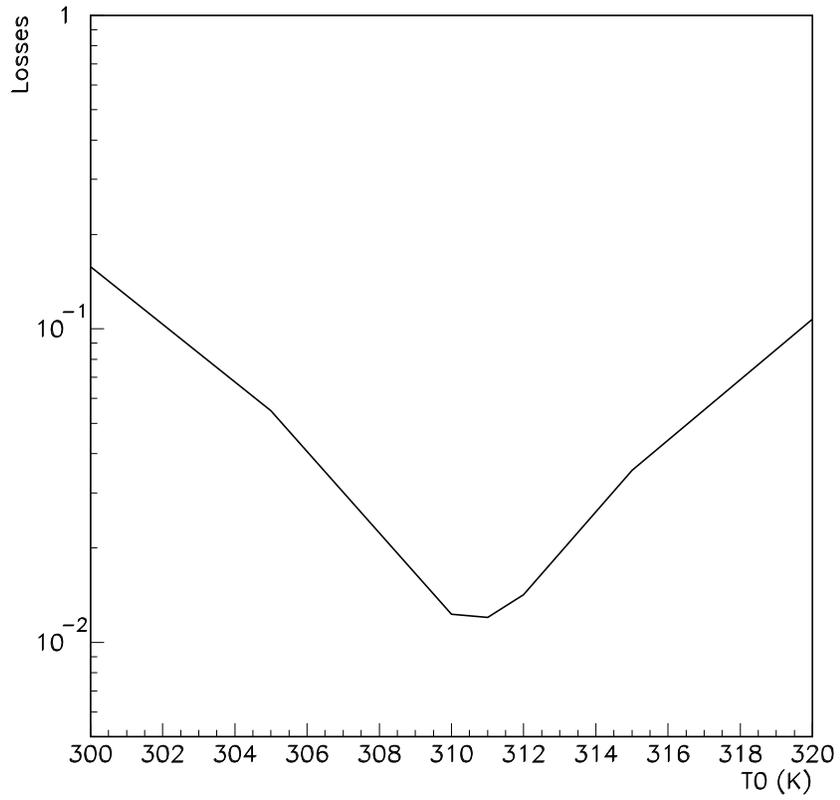,height=12cm}} 
\caption{Losses a as a function of the thermostat temperature. The tower temperature is fixed to
$T_e \simeq 300$ K.}
\end{figure}
\begin{figure}[H]
\centerline{\epsfig{file=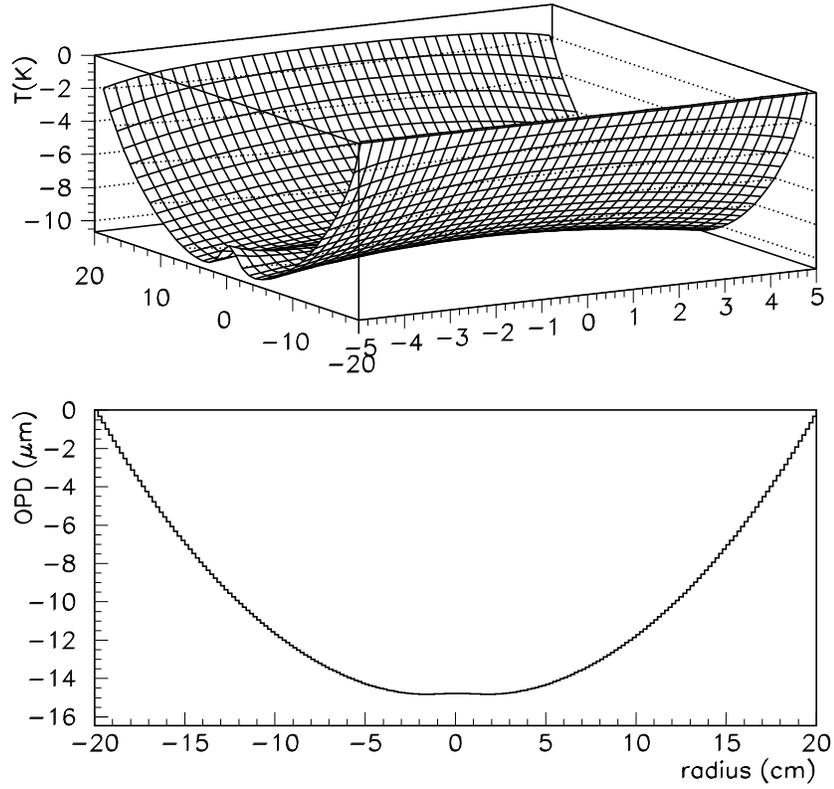,height=12cm}} 
\caption{Top: temperature field in the mirror. Bottom: thermal lens profile. Parameters: $T_e=300$ K
and $T_0=311$ K (optimum case). }
\end{figure}
\begin{figure}[H]
\centerline{\epsfig{file=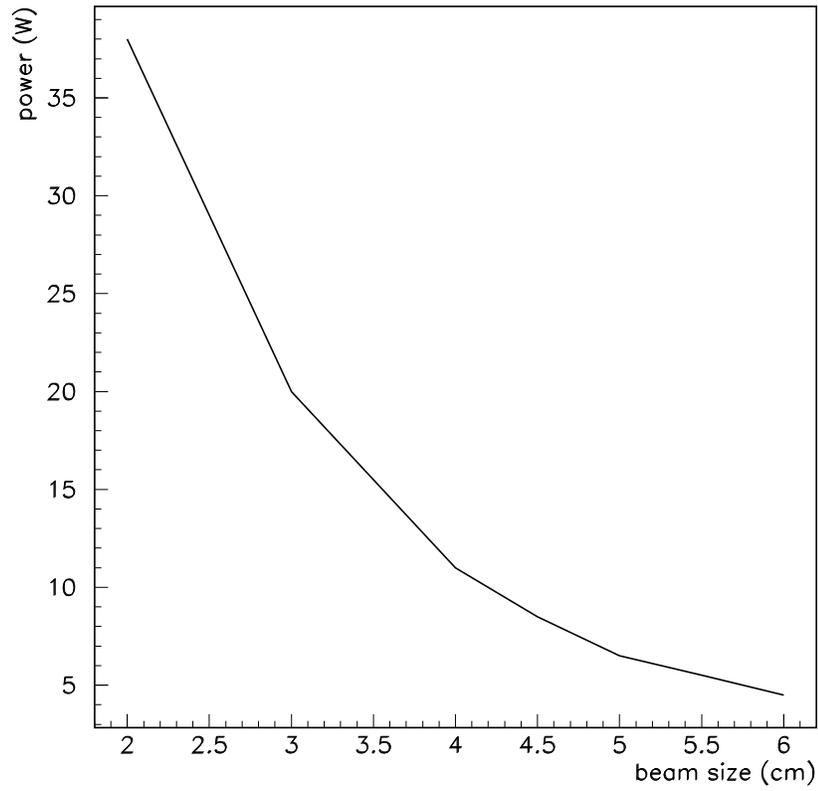,height=12cm}} 
\caption{Power dissipated in an extra ring-shaped beam in order to obtain mimimal losses, as a function
of the (main) beam size. The minimal losses are in each case about $2\times 10^{-3}$.}
\end{figure}
\begin{figure}[H]
\centerline{\epsfig{file=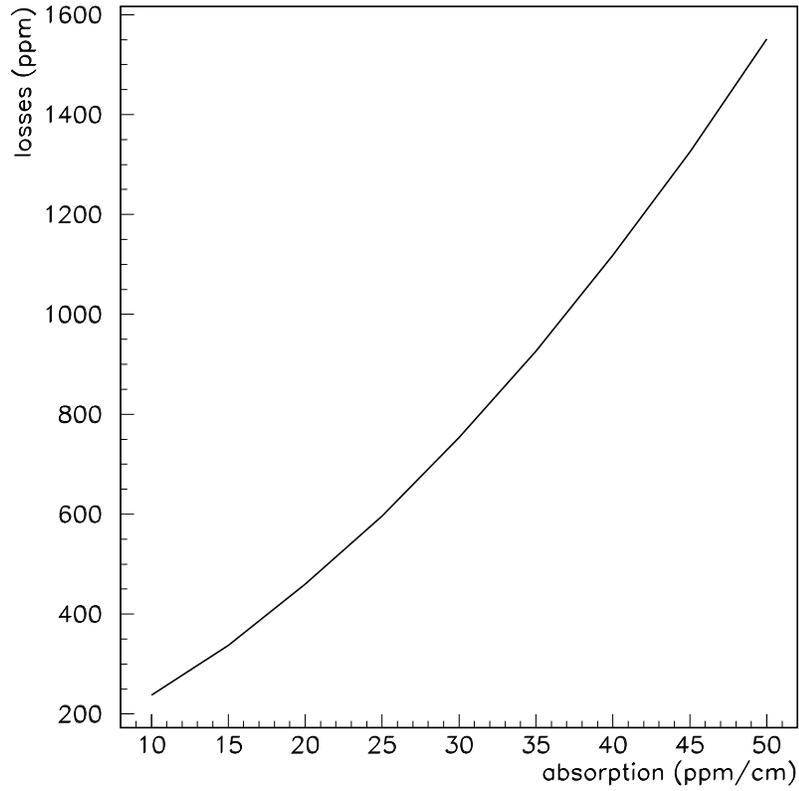,height=12cm}} 
\caption{Losses (in ppm) vs the Sapphire substrate absorption. For each absorption, the losses are optimized
with respect to the inner radius of the compensation beam and to the absorbed power from this beam.}
\end{figure}

\end{document}